\documentclass[12pt,a4paper]{article}

\usepackage[cp1251]{inputenc}

\usepackage[english]{babel}

\newtheorem{theorem}{Theorem}{\bf}{\it }
{\bf}{\it }
{\bf}{\it }
{\bf}{\it }
{\bf}{\it }
{\bf}{\it }

\title{Encipher of information on the basis of geometrical presentations}

\author{A.S. Petrov, A.D. Plotnikov\thanks{Department "Computing systems and networks"
East-Ukrainian national university  named for  V. Dal. 
}}

\date{}

\begin{document}
\maketitle

\begin{abstract}
In this paper, we examine a ciphertext on the basis of using geometrical objects. Each 
symbol normative alphabet is determined as a point on the plane. We consider possible 
ways for presentation of these points.
\end{abstract}



\section{Statement of the problem}

Let there is some alphabet $A=\{a_{1},a_{2},\ldots,a_{L}\}$ and there is a plaintext 
\[
M=(m_{1},m_{2},\ldots,m_{N}),
\]
where $m_{k}\in A$ ($k=1,2,\ldots,N$).

The alphabet $A$ is called {\it normative}, and symbols (letters) of the normative alphabet is presented  
by natural numbers  1,2, \ldots, $L$. As a rule, these numbers are represented in the 
binary scale of notation.

Formally, the plaintext $M$ is the ordered excerpts with repetition of elements of $L$-set $A$. 
The number of such excerpts \cite{1}:
\[
L^{N}
\]
and this number determines complexity of  random writing $M$ of symbols of the normative alphabet $A$.

For the information hiding, the plaintext $M$ is enciphered by means of the key $K$ and, as a result, the source text 
is transformed into a ciphertext $C=\phi(M,K)$. 

We can assume that any symmetric cruptosystem can be examined as a multipole, which transforms $N$ Boolean  
variables $m_{k}\in \{0;1\}$ into $N$  Boolean functions $c_{k}\in \{0;1\}$. Transformation of data on every iteration 
of encryption can be presented by a system of Boolean functions. To impede opening the key, this Boolean functions 
must be {\it nonlinear}. This is reached by embedding so-called $S$-boxes.

Besides, to increase security of the symmetric cryptosystems, key length is increased. However, processing power 
of computers grows steadily, therefore, key length grows also. For example, key length of cryptosystem AES can 
be 256 bits, and Blowfish can be 448 bits. This competition between the processing power and key length can complicate 
a cryptosystem considerably.

Aim of this paper be to examine organization of non-linearity in the process of encryption due to 
presentation of the plaintext by means of the geometrical objects. This presentation can be used with any existing 
symmetric cryptosystem or instead of $S$-boxes.

\section{Some variants of the use of geometrical objects}
 
We will notice, as a value of the symbol $m_{k}\in M$ is determined by  a value of a function $f(k,i)$, then plaintext 
$M$ can be given by points on the plane $k0i$. Clear, what such presentation is a 
graphic interpretation of $M$.

We will consider such possibilities.

Let a normative alphabet be symbols (letters) of the Roman alphabet, which correspond the   
followings natural numbers: 

\begin{center}
\begin{tabular}{|p{2.0cm}|c|c|c|c|c|c|c|c|c|}
\hline
Normative & & & & & & & & & \\
alphabet & A & B & C & D & E & F & G & H & I \\
\hline
Numerical & & & & & & & & & \\
equivalents & 1 & 2 & 3 & 4 & 5 & 6 & 7 & 8 & 9  \\
\hline
Normative & & & & & & & & & \\
alphabet & J & K & L & M & N & O & P & Q & R \\
\hline
Numerical & & & & & & & & &  \\
equivalents & 10 & 11 & 12 & 13 & 14 & 15 & 16 & 17 & 18  \\
\hline
Normative & & & & & & & & & \\
alphabet & S & T & U & V & W & X & Y & Z & \_ \\
\hline
Numerical & & & & & & & & & \\
equivalents & 19 & 20 & 21 & 22 & 23 & 24 & 25 & 26 & 27\\
\hline
\end{tabular}
\end{center} 

We will notice that the numerical equivalents do not contain a zero. It is important for 
further. 

Then the phrase "I\_LOVE\_MY\_MOTHER" can be presented the followings pairs:
\[
(1,9), (2,27),(3,12), (4,15), (5,22),(6,5),(7,27), (8,13),
\]
\begin{equation}
\label{ya}
(9,25),(10,27),(11,13),(12,15),(13,20),(14,8),(15,5),(16,18).
\end{equation}

Clearly we can omit of first the components of the pairs and can consider as a sequence of 
second components of the ordered pair: 
\begin{equation}
\label{ya1}
(9,27,12,15,22,5,27,13,25,27,13,15,20,8,5,18).
\end{equation}

The obtained co-ordinates of two-dimensional space can be presented as intersections some 
geometrical lines. In simplest case --- as intersections lines on a plane. In future, 
for illustration of results of the use of geometrical approach, we will use, mainly, the straight 
lines. We will consider an example. 

We suppose that the required lines pass through nearby points in the obtained sequence. The first and 
last points are also considered nearby. As is generally known, the normative equation of a straight line has 
form: $y=ax+b$ and is determined coefficients $a$ and $b$ fully. We will find these coefficients of a straight lines in our 
example. 

Equation of a straight line, passing through two points ($x_{1},y_{1}$) and ($x_{2},y_{2}$), looks like
\[
\frac{y-y_{1}}{y_{2}-y_{1}}=\frac{x-x_{1}}{x_{2}-x_{1}}
\]
or
\[
y=\frac{y_{2}-y_{1}}{x_{2}-x_{1}}x-\frac{y_{2}-y_{1}}{x_{2}-x_{1}}x_{1}+y_{1}.
\]
Here, $y\not=0$ in our case (see above).

For the first pair of points (1,9) and (16,18) we have:
\[
y=0.6x+8.4,
\]
that is, here $a=0.6$, $b=8.4$.

For the next pair of points (1,9) and (2,27):
\[
y=18x-10,
\]
that is, $a=1$, $b=29$.

Continuing, we will obtain the followings results:

\begin{center}
\begin{tabular}{|c|c|c|c|c|c|c|c|c|c|c|c|c|c|c|}
\hline
Pair of & (1,9) & (2,27) & (3,12) & (4,15) & (5,22) & (6,5) & (7,27) & (8,13) \\
points & (2,27) & (3,12) & (4,15) & (5,22) & (6,5) & (7,27) & (8,13) & (9,25) \\
\hline 
a & 18.00 & -15.00 & 3.00 & 7.00 & -17.00 & 22.00 & -14.00 & 12.00 \\
\hline
b & -9.00 & 57.00 & 3.00 & -13.00 & 107.00 & -127.00 & 125.00 & -83 \\
\hline
Pair of & (9,25) & (10,27) & (11,13) & (12,15) & (13,20) & (14,8) & (15,5) & (16,18) \\
points & (10,27) & (11,13) & (12,15) & (13,20) & (14,8) & (15,5) & (16,18) & (1,9) \\
\hline 
a & 2.00 & -14.00 & 2.00 & 5.00 & -12.00 & -3.00 & 13.00 & 0.60 \\
\hline
b & 7.00 & 167.00 & -9.00 & -45.00 & 176.00 & 50.00 & -190 & 8.40) \\
\hline
\end{tabular}
\end{center}

Finding intersections nearby straight lines, we will restoration the initial plaintext. 

However, the presentation of the initial plaintext by coefficients of the intersecting straight lines requires
passing to redundant information. To pass the plaintext, it is necessary to pass two identical arrays of data, 
one of which contains the values of coefficients $a$, and other --- $b$. At the described approach we 
must send coefficients for the straight lines, the number of which is equal to the common 
number of symbols in the initial plaintext. That is, the number of the straight lines is equal $N$ --- to length 
of the plaintext $M$. 

It will obtain an analogical result if instead of the straight lines we will be 
use an ellipse, also passing through two given points. Equation of such ellipse looks like: 
\[
y^{2}=x^{3}+ax+b.
\]
In this case, all said above is valid, but the formulas of calculation for ellipses are more complexity.

To decrease length of arrays, containing the values of coefficients $a$ and $b$, it is possible, if symbols of the initial 
plaintext split on pairs. If the initial plaintext has odd length, we can add symbol of  interval in the end or beginning 
the plaintext. 

Because the phrase "I\_LOVE\_MY\_MOTHER", in which symbols of the normative alphabet are transferable by their numerical 
equivalents, looks like: (9,27,12,15,22,5,27,13,25,27,13,15,20,8,5,18). Then the partition on pairs looks so: 
\begin{equation}
\label{c1}
((9,27),(12,15),(22,5),(27,13),(25,27),(13,15),(20,8),(5,18))
\end{equation}

Examining every pair in the partition (\ref{c1}) as a co-ordinates of point on a plane, one can present 
these points as intersections of the straight lines, passing through nearby points. As well as in the previous example 
we suppose that the first and last points determine a straight line also. Now, we must give  
information about straight lines, the number of which is half less, than in the previous case.

However, the equation of a straight line, passing, for example, through the pair of points (12,1) and (12,19), is $x=12$, but it  
can not be presented in the form $y=ax+b$. By the way, application of an elliptic curve ($y^{2}=x^{3}+ax+b$) 
leads to the same result.

The indicated difficulty can be overcame, supposing that every equation of a straight line, connecting nearby points, 
it is presented in a general form: 
\begin{equation}
\label{eq1}
Ax+By+C=0.
\end{equation}

Then formulas for the calculation of coefficients of general equation of the line (\ref{eq1}), passing through two 
set points $(x_{1},y_{1})$ and $(x_{2},y_{2})$, look like:
\[
A=y_{2}-y_{1};\ \ \ B=x_{1}-x_{2};\ \ \ C=y_{1}(x_{2}-x_{1})-x_{1}(y_{2}-y_{1}).
\]

For our plaintext "I\_LOVE\_MOTHER", the coefficients of intersecting straight lines are presented in the following 
table.  

\begin{center}
\begin{tabular}{|c|c|c|c|c|c|c|c|c|c|c|c|c|c|c|}
\hline
Pairs of & (9,27) & (12,15) & (22,5) & (27,13) & (25,27) & (13,15) & (20,8) & (5,18)\\
points & (12,15) & (22,5) & (27,13) & (25,27) & (13,15) & (20,8) & (5,18) & (9,27)\\
\hline 
A & -12    & -10 & 8        & 14    & -12 & -7     & 10 & 9\\
\hline
B & -3      & -10 & -5       & 2      & 12  & -7     & 15 & -4\\
\hline
C &  189  & 270 &  -151 & -404 & -24  & 196 & -320 & 27 \\       
\hline
\end{tabular}
\end{center}

In this approach, the information about the intersecting straight lines can be passed by an arrays of coefficients, length each of which 
is half of length of the initial plaintext. 

In common case, we can split symbols of the initial plaintext into groups from $p$ symbols and each such group we 
examine as a point of $p$-dimensional space.  

\begin{theorem}
The most dimension of space, in which it is possible to examine an initial plaintext, is equal $p=[N/3]$, 
where $[a]$ is the least integer, exceeding $a$.
\end{theorem}

To use the offered approach, it is necessary, obviously, to pass information at least three 
points. \hfill Q.E.D. 

\vspace{0.5pc}
We will consider another approach to presentation of information, using the approximation theory of functions. We can present 
points of space, corresponding the initial plaintext, by the polynomial of Lagrange. Because in this case, the polynomial has  
a degree $n$, where $n$ is the number of points, it is expedient the initial plaintext split on groups, for example, for four points. In this case 
the plaintext will be presented next tables, corresponding each of groups. 

The combined table for these groups looks like:

\vspace{0.5pc}
\begin{center}
\begin{tabular}{|c|c|c|c|c||c|c|c|c||c|c|c|c||c|c|c|c||}
\hline
$i$ & 0 & 1 & 2 & 3 & 0 & 1 & 2 & 3 & 0 & 1 & 2 & 3 & 0 & 1 & 2 & 3 \\
\hline
$x_{i}$ & 1 & 2 & 3 & 4  & 5 & 6 & 7 & 8 & 9 & 10 & 11 & 12 & 13 & 14 & 15 & 16\\
\hline
$y_{i}$ & 9 & 27 & 12 & 15 & 22 & 5 & 27 & 13 & 15 & 20 & 8 & 5 & 18 & 27 & 27 & 27 \\
\hline
\end{tabular}
\end{center}
\vspace{0.5pc}

From the first table we obtain a polynomial:
\[
P_{1}(x)=-93+161x-67.5x^{2}+8.5x^{3}.
\]
From the second table we have:
\[
P_{2}(x)=184.2-207.72x+48.7x^{2}-3.18x^{3}.
\]
The third table gives a polynomial:
\[
P_{3}=-55.67+82.91x-12.77x^{2}+0.52x^{3}.
\]
A fourth table gives a polynomial:
\[
P_{4}=15.92+2.22x-0.15x^{2}+0.003x^{3}.
\]

Thus, instead of the sequence (\ref{ya1}) we have a sequence
\[
(-93;161;-67.5;8.5;184.2;-207.72;48.7;-3.18;
\]
\[
-55.67;82.91;-12.77;0.52;15.92;2.22;-0.15;0.003).
\]

It is not difficult verify that for the calculation of true sequence (\ref{c1}) on the built polynomials, it is necessary 
to find the integer values of co-ordinates of the corresponding points on the rules of rounding off. 

It seems that similar approach, apparently, is not effective.

\section{Conclusion}

As follows from the considered variants of application of geometrical presentations for enciphering of texts, 
merit of this method consists of difficulty of analysis of ciphertext, because the use is practically impossible 
Boolean functions for description of procedure of enciphering declaration. Lack consists of that it requires excessiveness 
at the use. Clear, that such approach requires additional researches.

\end{document}